\documentclass[review]{elsarticle}

\usepackage{lineno,hyperref}
\modulolinenumbers[5]

\journal{Journal of \LaTeX\ Templates}

\begin{document}

\begin{frontmatter}

\title{Cosmic electron spectra by the Voyager instruments and the Galactic electrostatic field}

\author{Antonio Codino\fnref{myfootnote}}
\address{University of Perugia, and Scientific associate of $INFN$, Italy}
\fntext[myfootnote]{antonio.codino@pg.infn.it}

\begin{abstract}
The Voyager spacecrafts have been measuring since 2012 the rates of electron
       and nuclei of the cosmic radiation beyond the solar cavity
       at a distance of more than $10^{13}$ $meters$ from the Earth. A record of unique
       and notable findings have been reported and, among them, the
       electron-to-proton flux ratio of 50 to 100 below energies of $50$ $MeV$.
This ratio is thoroughly opposite of that of 0.01 measured at higher
energies in the range 10 $GeV$ to 10 $TeV$. The difference amounts
to four orders of magnitude. Arguments and calculations to show how
this surprising and fundamental ratio lends support to the empirical
evidence of the ubiquitous electrostatic field in the Milky Way
Galaxy are presented. In other respects this paper examines and
calculates, for the first time, the electric charge balance in the
solar system delimited by the $termination$ $shock$ of the solar
wind.
\end{abstract}

\begin{keyword}
\texttt{cosmic rays in the Sun, solar modulation, Sun and stellar magnetic fields, electric fields in the solar system}
\end{keyword}

\end{frontmatter}


\section{Introduction}
 A recent work provides empirical evidence of the Galactic
electrostatic field \cite{cod20ubibook}.  A very short account of this work has been
presented at the 37-th ICRC 2021, Berlin, Germany \cite{cod22ubi}. The Galactic
electric field designated by $\vec E_g$ (g for galactic) is
permanent and ubiquitous except in regions where electrostatic
shielding of structured ionized materials occurs. The solar system
is one of this shielded region and, in spite of the electric
shielding, highly specific effects with unmistakable signatures
manifest themselves attesting the existence of the Galactic electric
field $\vec E_g$. This paper deals with one of these effects, that
is, the energy spectrum of cosmic electrons at low energies below
$80$ $MeV$ and above $3$ $MeV$ measured by the Voyager spacecraft.

\quad In essential terms and in other context, the negative electric
charge of the electron spectrum, $3$-$80$ $MeV$, is just the
electric charge needed to continuously neutralize the positive
electric charge deposited by cosmic rays in the solar cavity. Hence
this paper examines, calculates and reports, for the first time, the
electric charge balance in the solar system within the $termination$
$shock$ of the solar wind.

\quad A stuffy introduction is avoided here by relegating in the
$Appendix$ $A$ the new scientific context where the present paper
properly shines.

\par\parskip=0.5truecm \quad \quad According to the
research book \cite{cod22hoewelebook} published at the end 2022, \quad $How$
$electrostatic$ $fields$ $generated$ $by$ $cosmic$ $rays$ $cause$
$the$ $expansion$ $of$  $the$ $nearby$ $universe$,  \quad all stars
of the $Galaxy$ retain a notable amount of positive electric charge
deposited by the Galactic cosmic-ray nuclei. Charge deposition
occurs by two different ways in two separate regions : the first way
is the stopping of low-energy cosmic rays in the solar wind  and the
second way is the stopping of all cosmic rays of low and high
energies in the $Solar$ body by ionization and nuclear interactions.
By solar body is meant all the high density material residing within
photosphere of radius $R_s$ ($s$ for $Sun$). The electric charge per
second deposited in the $Sun$ will be designated by $I_{sw}$ ($s$
for solar $w$ for wind) and $I_{ds}$ ($d$ for dense and $s$ for
$Sun$), respectively. In the following the photospheric radius
$R_{s}$ of the  $Sun$ is $6.98 \times 10^{5}$ $km$ and the massk
inside $1.989 \times 10^{33}$ $g$ so that the average density is 1.4
$g$/$cm^3$. Anticipating successive results, above $5$ $GeV$,
$I_{ds}$ = $1.1 \times 10^{3}$  $Coulomb$/$s$  (hereafter $C$/$s$)
and $I_{sw}$ = $1.96 \times 10^{12}$ $C$/$s$  in the range $3$
$MeV$- $20$ $GeV$ of the cosmic-ray spectrum where data are
available.

 \quad The region occupied by the solar wind is approximated by a sphere
 of radius $R_{sc}$ (sc for solar cavity). Its volume,
  $V_{sc}$ $\equiv$ (4/3) $\pi$
$R_{sc}^3$ \quad is a surrogate of the non spherical volume
delimited by the termination shocks of nominal radius $R_{sc}$ of
$1.331 \times 10^{13}$ $m$ observed by the Voyager Probes in two
different positions  of the nearby circumstellar space \cite{stone05,richa08}. The
two positions in heliocentric system are $37.5$ degrees North at 94
AU \cite{cod22hoewelebook} (Astronomical Unit, $1.49597870 \times 10^8$ m) and -$116.5$
degrees South at $83.7$ $A$$U$ \cite{stone05}. Crudely and nominally, $R_{sc}$
$\equiv$ (94 + 84)/2 = $89$ $A$$U$ = $1.331 \times 10^{13}$ $m$ and
$V_{sc}$ = ($4$/$3$) $\pi$ $R_{sc}^3$ = $9.8769 \times 10^{39}$
$m^3$. In this work solar cavity is synonimous of solar cavern.\\ 

\quad  The electric charge deposition in the $Sun$ by cosmic nuclei
mentioned above also apply to any stars as stellar winds are
universal (see, for example, \cite{kudritzki2000}).

\section{Charge deposition in the solar wind region}

\quad  Nuclei and electrons of the Galactic cosmic radiation
intercepting the solar wind, in $Classical$ $Mechanics$ a sort of
continuous fluid moving outwards, loose energy and a conspicuous
fraction of the cosmic-ray flux with energies below $1$-$10$
$GeV$/$u$ is thoroughly arrested  and dispersed in the environment.

\quad An artistic portrait of the electric charge deposited by
cosmic rays in the solar wind is shown in fig. 1 by red crosses.

\quad Let $J_{T}$  ($T$ for terrestrial flux)  denote the flux of
cosmic rays measured at $Earth$ at the top of the atmosphere (zero
atmospheric depth) and $J_{de}$ ($d$$e$ for demodulated) the
unperturbed flux in the nearby interstellar space usually called
demodulated flux. Solar modulation cyclically affects cosmic-ray
intensity below $10$-$20$ $GeV$ within the solar wind volume (see,
for example, ref. \cite{usoskin11}). The total amount of electric charge, \quad
$Q_{sw}$ \quad deposited by cosmic rays in the volume $V_{sc}$
occupied by the solar wind $Q_{sw}$ is given by :
\begin{equation}
 I_{sw}  =   \overline{q}_{cr} ( E)  A_{sc} \int ^{E_{2}}_{E_{1}} {  J_{de}} dE
\end{equation}  
where $A_{sc}$ = $4$ $\pi$ $R_{sc}^2$ = $2.2262 \times
10^{27}$ $m^2$ is the area intercepted by the demodulated flux of
cosmic rays in the energy band $E_1$ and $E_2$ being  $E_2$ $\ge$
$E_1$. For protons $\overline q_{cr}$ = $1.0 \times 10^{-19}$ $C$
and for all cosmic nuclei at low energy $\overline q_{cr}$ = $1.2
\times 10^{-19}$ $C$ (see fig. 2 in ref. \cite{cod15icrc}). Of course, the electric
charge deposited between the $Earth$ radius and the $Sun$
photosphere is missed in equation (1) and it has to be summed up.
Here only notice that measurements of the cosmic-ray flux at
$0.1$-$1$  $AU$ performed by recent and past space missions (Parker
Solar Probe, Helios, Ulysses ) are available.

\quad The demodulated spectrum besides nuclei include electrons
$J_e$, positrons $J_{pos}$ and antiprotons $J_{\overline p}$ so that
the global positively charged cosmic-ray flux is, \quad $J_{de}$ -
$J_e$ + $J_{pos}$ - $J_{\overline p}$. Traditionally in the
literature the demodulated spectrum is also termed $Local$
$Interstellar$ $Spectrum$  (LIS)  but this is an interpretation of
the demodulated spectrum, and not pure data, like energy spectra
resulting from measurements  are.

\begin{figure*}
\begin{center} 
\includegraphics[width=0.4\textwidth]{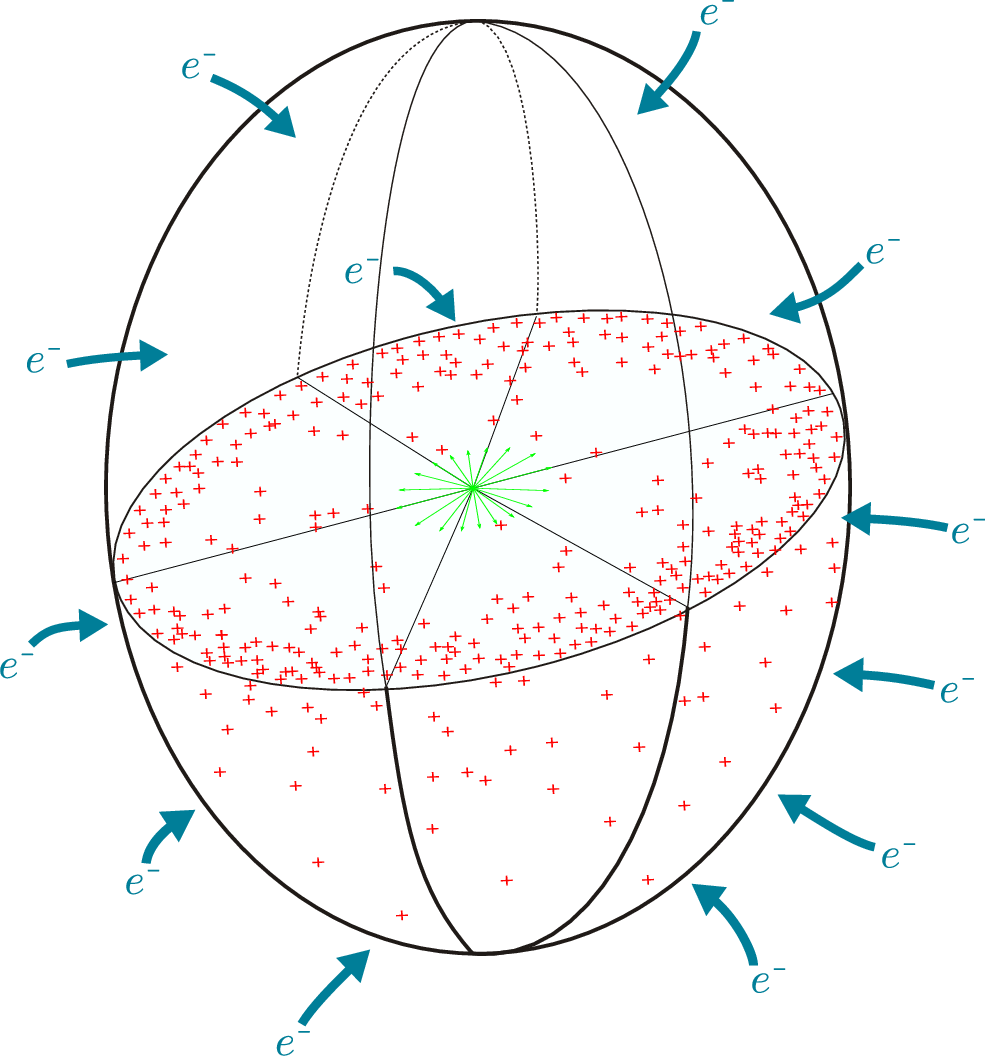}
\caption{Artistic portrait of low-energy cosmic rays
extinguished in the solar wind. The positive electric charge
deposited by cosmic rays extinguished in the solar wind  is marked
by red crosses. Solar wind stems from the corona and is propelled in
the corona. Quiescent electron currents (represented by thick blue
arrows) coming from the nearby circumstellar space migrate toward
the region occupied by the solar wind  where cosmic-ray extinction
takes place. Green vectors of fanlike shape represent the
electrostatic field $\vec E_s$ at the base of the $Solar$
photosphere projected in an arbitrary plane,  for simplicity. The
ovoidal volume of the figure is arbitrary. \label{fig:36trab}}
\end{center}
\end{figure*}

\begin{figure}
\begin{center} 
\includegraphics[width=0.95\textwidth]{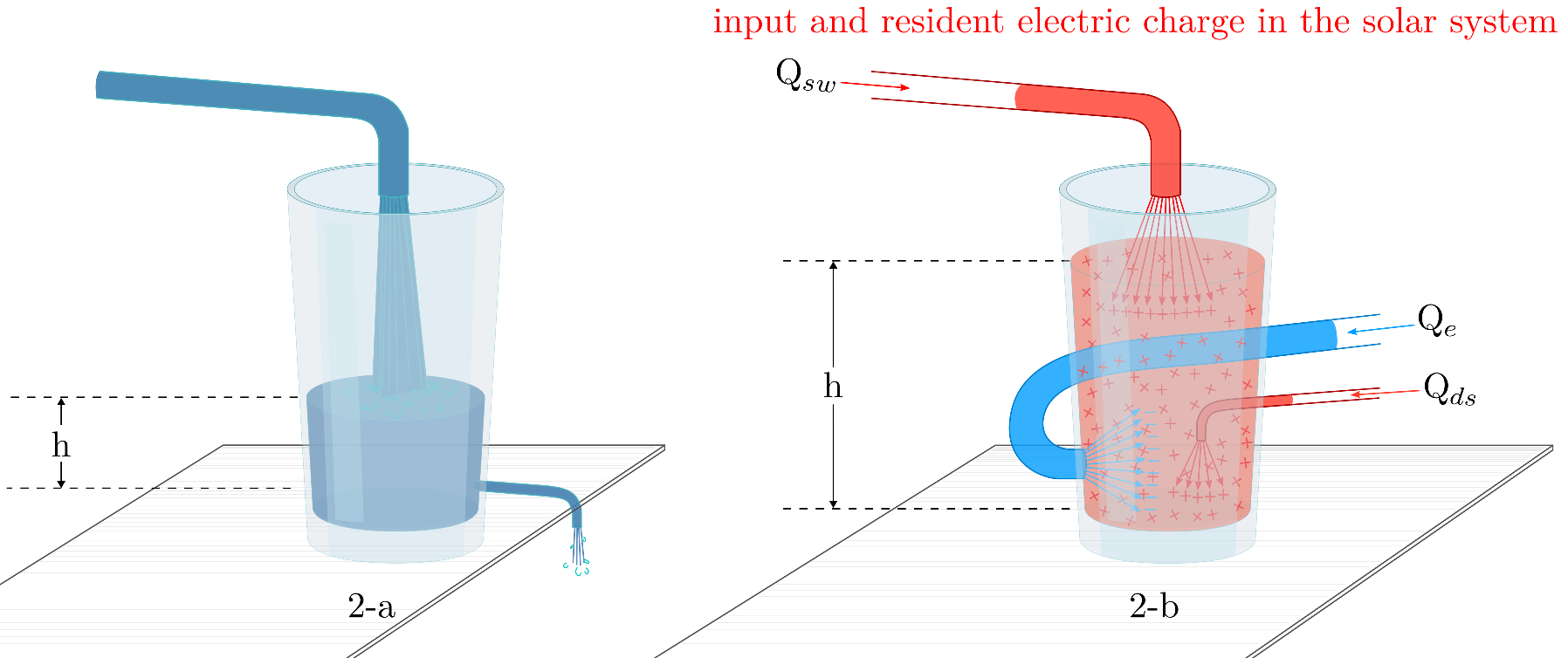}
\caption{Illustration of electric charge flows in the
solar cavity by a simple hydraulic analogy.  \quad (2-a) The gap $h$,
in the liquid levels in the bucket is not zero (an empty bucket
corresponds to $h$ = 0) but it has some finite value governed by
gravity pressure, input and output calibers and input velocity.
\quad (2-b) Red pipelines entering the bucket from top and side in
the arbitrary time interval $\delta$(t) represent the positive
charge \quad  $Q_{sw}$ + $Q_{ds}$ \quad which equals the negative
charge $Q^-_{ne}$ represented by the blue pipeline. Here the bucket
denotes the nominal volume of the solar cavity $V_{sc}$ = $9.8769
\times 10^{39}$ $m^3$. The stationary condition in the charge
balance within $V_{sc}$  is referred to as equilibrium or quasi
steady condition. Equilibrium implies that some residual positive
charge, $R$($\delta$$t$) defined by equation (2), permanently
remains in the bucket in a regime where positive and negative
currents \quad $Q_{sw}$ + $Q_{ds}$ + $Q^-_{ne}$  \quad balance. From
this charge, namely $R$($\delta$t),  originates the permanent
electrostatic fields $E_{sw}$ and $E_{ds}$. \label{fig:voy-2}}
\end{center}
\end{figure}

\begin{figure}
\begin{center} 
\includegraphics[width=0.8\textwidth]{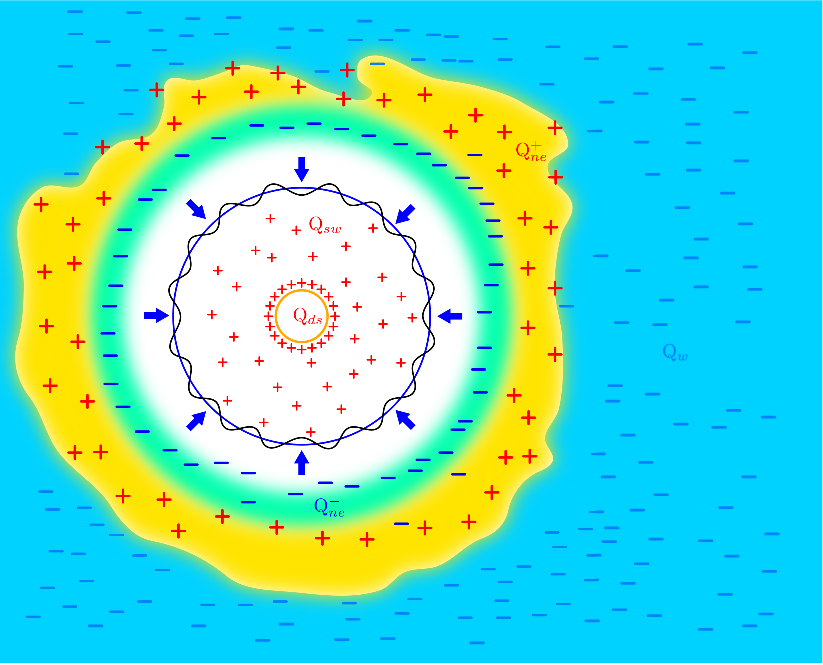}
\caption{Highly schematic view of spherical regions around
the Sun where positive and negative charges are expected to prevail.
The wavy black curve designates the termination shock of the solar
wind observed by the Voyager Probes $V1$ and $V2$ encircling the
positive electric charge of volume $V_{sw}$. The charge $Q^-_{ne}$
of the migration current occupies the spherical green region with
negative charge prevalence around the solar cavity where low energy
electrons are expected to dominate. The yellow region with positive
charge (plus signs) surrounding the green negative charge region
(minus signs) is depleted of low energy electrons migrated in the
inner adjacent green region surrounding the termination shock (wavy
black region). The vast blue region with negative charge $Q_w$ (blue
minus signs) designates the $widow$ $electron$ region. Widow
electrons are defined as orbital electrons abandoned in the
environment by cosmic nuclei during the acceleration stage at the
cosmic-ray sources. \label{fig:109}}
\end{center}
\end{figure}

\section{Charge balance around stars and multiple star systems}

\quad According to Potassium abundance in terrestrial meteorites
\cite{voshage83} the cosmic-ray intensity in the last 2 billion years has
maintained approximately constant. This datum is regarded as solid
input here to state,  or  posit,  the quasi stationary state  in the
flows of the electric charge in the volume $V_{sc}$.

\quad At the arbitrary time interval $\delta$$t$,  the positive
electric charge deposited in the volume $V_{sc}$ amounts to :
($I_{sw}$ + $I_{ds}$) $\delta$$t$.

\quad  The neutralization of the positive charge excess during the
time span $\delta$$t$, \quad $I_{sw}$$\delta$$t$ +
$I_{ds}$$\delta$$t$ \quad in the volume $V_{sc}$ promotes a
quiescent electron current designated by $I_e$ and called
\emph{migration current}. The electric current $I_e$ is
qualitatively represented by the thick blue arrows in fig. 1. The
current $I_e$ entering the solar cavity coming from the nearby
interstellar space, beyond the radius $R_{sc}$ ,  transports the
negative electric charge $Q_e$ = $I_e$$\delta$$t$ in the arbitrary
time interval $\delta$$t$. For an ideal, continuous isotropic
current and a spherical solar cavity it would be :
$Q_e$($\delta$$t$) = $4$ $\pi$ $R_{sc}^2$ $I_e$$\delta$$t$. The
total electric charge contained in the solar cavity of volume
$V_{sc}$ in the time interval $\delta$$t$ is given by :
\begin{equation}
 (I_{e} +   I_{sw}+  I_{ds} ) \delta t  + q_{sc} (0) =  R(\delta t)
\end{equation}
\quad where $q_{sc}$ (0)  is the total electric charge in the volume
$V_{sc}$ at initial time $t$ = 0. Any arbitrary time may be chosen
as initial time or zero time taking into account the $Sun$ age of
about $4.5673$ $Ga$. The charge within the volume $V_{sc}$ = $9.8769
\times 10^{39}$ $m^3$ at time $\delta$$t$ is denoted by $R$($\delta
t$).

\ The stationary state during the time interval $\delta$$t$ posits,
\quad  $I_{e}$ + $I_{sw}$ + $I_{ds}$ = $0$  \quad and, therefore,
$q_{sc}$(0) = $R$($\delta t$) where $R$($\delta t$) designates the
residual electric charge at the time $\delta t$. For instance,
during the temporal span surveyed by Potassium data \cite{voshage83} of
$\approx$ $2$ $Ga$, the residual charge $R$($ 2 Ga$) remained
constant and the time $t$ = $0$, i.e. the initial time of the
present calculation is $2$ $Ga$ back in time from the present epoch.

 \quad  If the average displacement velocity of the migration current $I_e$
 were very high, ultimately,
comparable
 to $c$, the light speed, a rapid neutralization
of the positive electric charge ($Q_{sw}$ + $Q_{ds}$) = ($I_{sw}$ +
$I_{ds}$) $\delta t$  within the volume $V_{sc}$ at any arbitrary
time spans  $\delta$t would take place. In this imaginary and
unphysical condition the residual charge $R$($\delta t$) in equation
(2) would turn out to be $q_{sc}$(0). The neutralization of the
positive electric charge, \quad $I_{sw}$$\delta t$ + $I_{ds}$$\delta
t$ \quad requires a characteristic time interval $T_{ne}$ ($n$$e$
per neutralization) regulated by the conductivity of the
circumstellar medium, the distance of the cosmic-ray sources, the
convective motion of local materials, moving charged plasma parcels,
moving neutral plasma pockets in the environment, the local magnetic
fields, local electrostatic fields, the electrostatic field of the
star surfaces $\vec E_s$ and in the solar wind region $\vec E_{sw}$
and other parameters. During the time span $T_{ne}$ the negative
charge $Q_{ne}^-$ transported by quiescent electrons surrounding
nearby stars and nearby clouds (see fig. 3) uncovers the positive
quiescent charge $Q_{ne}^+$ in a larger circumstellar ambient so
that : $Q_{ne}^-$ + $Q_{ne}^+$ = 0, \quad based on charge
conservation in a finite volume of the adjacent larger ambient
surrounding the tiny solar cavity.

\quad  How much electric charge did the nascent $Sun$ store before
emanating solar wind ?  If the time $t$ = 0 is set at 4.567 $Ga$
back in time (nominal $Sun$ age), then the  term  $q_{sc}$($0$) in
equation (2) would represent the pristine, presolar electric charge
at the $Sun$ birth.

\quad Due to the finite neutralization times $T_{ne}$($\delta$$t$),
the positive electric charge $q_{sc}$($t$) accumulates implying
$q_{sc}$($t$) $>$ $q_{sc}$($0$). Ultimately, $q_{sc}$($t$) will
be neutralized by a fraction of the negative electric charge $Q_{w}$
= $2.58 \times 10^{31}$ $C$ dispersed in the cosmic-ray sources in
the whole Galactic disk. This negative charge are orbital electrons
lost by the accelerated cosmic nuclei in all Galactic cosmic-ray
sources. These particular orbital electrons are designated in the
twin works \cite{cod20ubibook,cod22hoewelebook} by the new term, $widow$ $electrons$. The term
is useful whenever charge balance of various particle populations
including cosmic rays has to be examined and calculated. This point
is debated in $Chapter$ $12$ of ref. \cite{cod20ubibook}.

\quad Taking into account the total negative charge of widow
electrons $Q_w$ and the positive charge  of  cosmic-ray nuclei
$Q_{cr}$ being $Q_{cr}$ = -$Q_w$ = $2.58 \times 10^{31}$ $C$,  a
more rigorous expression of equation (2) may be written in this way
:

$$ (I_{e} +   I_{sw}+  I_{ds} )
 \delta t  + Q_w(V_{sc}) + Q_{cr}(V_{sc}) + q_{sc} (0) =  R(\delta
 t)
    \eqno (2-b)$$

\quad  where $Q_w(V_{sc})$ is the nominal negative charge of widow
electrons in the volume $V_{sc}$ and $Q_{cr}(V_{sc})$ the nominal
positive charge of cosmic-ray nuclei in the same volume $V_{sc}$.
Therefore, $a$ $priori$ the nominal negative charge of $widow$
$electrons$ would be $Q_w$($V_{sc}$) = - $4.6 \times 10^{11}$ $C$
and + $Q_{cr}$($V_{sc}$) less than this value because the positive
charge is dispersed in a volume larger than that of the disk (see
next Section $4$).

\quad Notice that sources of electric charge within the solar cavity
do not alter the stationary state envisaged above. For example,
Jupiter planet releases electrons \cite{lheureux76,moses87} observed at $1$ $AU$ by
$ISEE$ $3$ at a rate of some $4.6 \pm  0.19 \times 10^4$
electron/$GeV$ $m^2$ $s$ $sr$ in the range $1$-$30$ $MeV$ \cite{moses87}.
Charge conservation requires a corresponding positive electric
charge in situ in the Jupiter planet. Other known sources of
negative charge within $V_{sc}$ are sporadic emission of energetic
electrons during solar flares.

\section{The equilibrium of the electric currents in the heliosphere}

\quad The electric charges $I_{ds}$$\delta$t and $I_{sw}$$\delta$t
stored in the $Sun$ and solar wind, respectively, during the time
span $\delta$t  produce distinctive physical effects. For example,
as the charge $Q_{ds}$ rotates with the $Sun$, it has to generate a
magnetic field. This effect is presented in $Section$  6.

\par The basic condition of the charge balance expressed by equation
(2) is the stationary state or steady condition or equilibrium. In
this condition, in the volume $V_{sc}$ = $9.87 \times 10^{45}$
$cm^3$, electric currents equilibrate e.g. \quad $I_{e}$ + $I_{ds}$
+ $I_{sw}$ = $0$ and the baseline charge, \quad $R$($\delta$t) =
$Q_w$($V_{sc}$) + $Q_{cr}$($V_{sc}$) + $q_{sc}$($0$) \quad is not
zero but it assumes some finite value as depicted in fig. 2-a by an
hydraulic analogy. In the steady-state condition at the arbitrary
time span $\delta t$, regardless of the complexity of the
environment, the neutralization charge $Q^-_{ne}$ $\equiv$ $I_e$
$\delta t$ entering the solar cavity from the adjacent interstellar
space (see fig. 3) has to equalize the positive electric charge
\quad ($I_{sw}$ + $I_{ds}$) $\delta t$ \quad deposited by cosmic
nuclei in the same volume and same time span $\delta t$. This
neutralization is certainly influenced by the permanent charge
$R$($\delta t$). The equilibrium implies that the number of
quiescent electrons of extremely low energy entering the solar
cavity has to swamp that of quiescent protons and quiescent nuclei
in the same energy band because of the overwhelming positive charge
entering the solar cavity via cosmic nuclei of higher energies,
above $60$ $MeV$ as attested in fig. 4.

\begin{figure}
\begin{center} 
\includegraphics[width=0.8\textwidth]{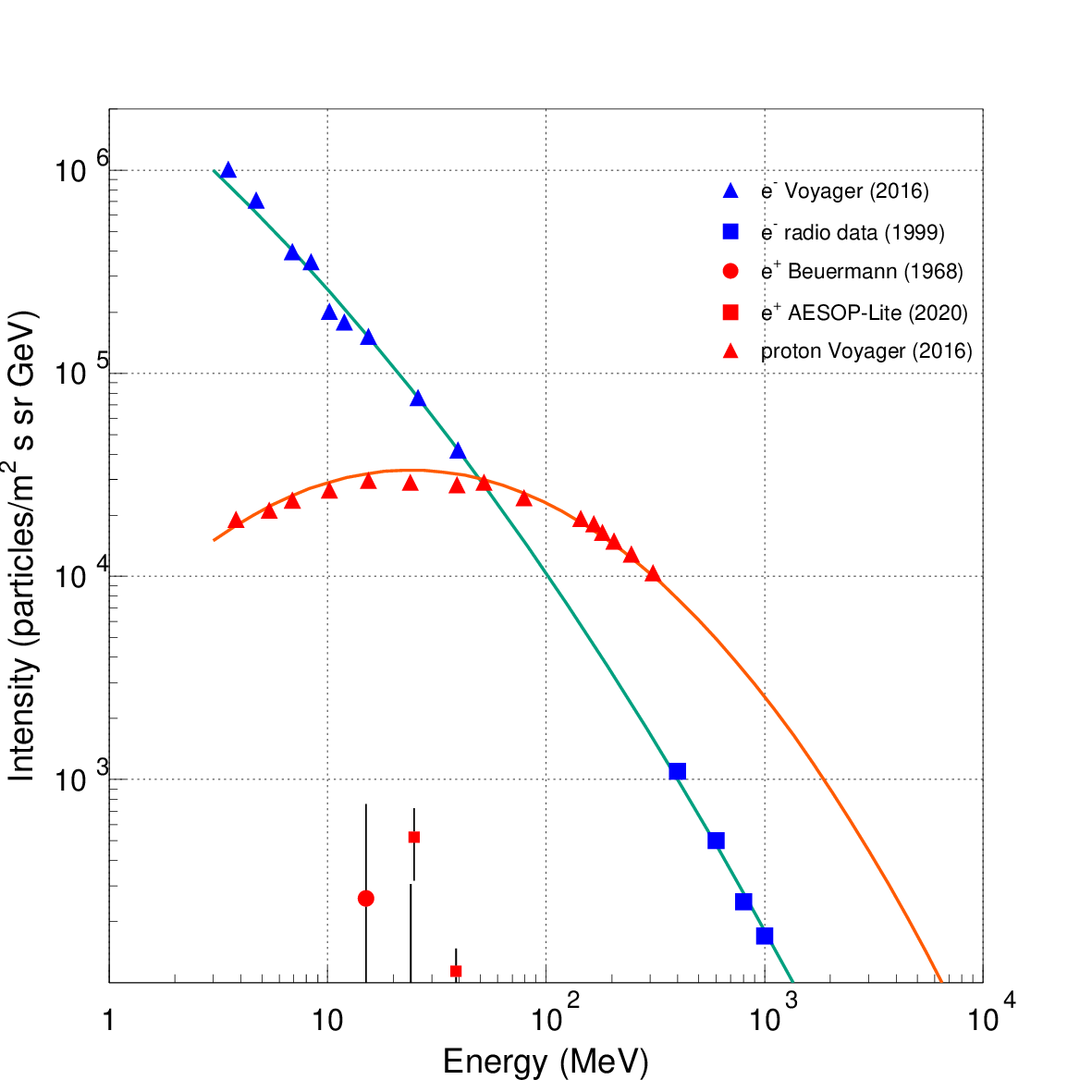}
\caption{Cosmic-ray electron spectrum measured by Voyager
in the range 2 $MeV$ to $100$ $MeV$  along with cosmic-ray proton in
the range $3$ $MeV$ to $600$ $MeV$ measured by the $Voyager$
$Probes$ \cite{cummings16}. The continuous blue line is the extrapolated solar
modulation curve. The continuous red line is the proton solar
modulation curve. Positron data at these low energies are scarce ;
positron intensity is are more than an order of magnitude below the
proton curve. \label{fig:110}}
\end{center}
\end{figure}

\quad  Notice that the nominal charge of widow electrons in the
solar cavity, $Q_{w}$($V_{sc}$), is thoroughly negligible compared
to the charge \quad $I_{sw}$$\delta t$ + $I_{ds}$$\delta t$ \quad
 deposited by cosmic nuclei. As the average
charge density of widow electrons in the Galactic disk of volume
$V_d$ is, $Q_w$/$V_d$ = -$2.58 \times 10^{31}$ $C$ /$5.19 \times
10^{66}$ $cm^3$ = -$0.498 \times 10^{-35}$ $C$/$cm^3$ \cite{cod20ubibook} and the
volume of the spherical solar cavity is $V_{sc}$ = $9.87 \times
10^{45}$ $cm^3$, nominally, the total negative charge
$Q_{w}$($V_{sc}$/$V_d$) is about $4.9 \times 10^{10}$ $C$,
thoroughly negligible relative to any plausible estimate of $I_{ds}$
$\delta$$t$  and $I_{sw}$$\delta$$t$. For example, by setting the
lifetime of cosmic rays in the Galaxy, $\delta$$t$ = $15 \times
10^6$ years = $4.73 \times 10^{14}$ $s$, it results $Q_{ds}$ =
$I_{ds}$ $\delta$$t$ = $1.1 \times 10^3$ $\times$ $4.73 \times
10^{14}$ = $5.20 \times 10^{17}$ $C$ and $Q_{sw}$ = $I_{sw}$
$\delta$$t$ = $1.96 \times 10^{12}$ $\times$ $4.73 \times 10^{14}$ =
$9.27 \times 10^{26}$ $C$, respectively, which outnumber
$Q_w$($V_{sc}$) limited to $4.9 \times 10^{10}$ $C$.

\quad As local electric fields exist in interstellar clouds as also
pointed out by

\quad others and the solar system is embedded in the nearby clouds
\samepage{\footnote{\it The volume of the space adjacent to the
solar cavity is insignificant relative to that of the $Galaxy$ of
some $2.55 \times 10^{66}$ $cm^3$. The $Local$ $Bubble$ surrounding
the $Sun$ has a size of about $10$ $pc$, gas density $0.05$
$atoms$/$cm^3$ and temperature of $10000$ $degrees$. The Local
Interstellar Cloud embraces the $Local$ $Bubble$ in a wider ambient,
about $300$ $pc$ in size,  and other bubbles called $Loop$ $1$,
$Loop$ $2$ and $Loop$ $3$ lie in the vicinity. The precious, unique
and historical results of the $Voyager$ $Probes$ on magnetic fields,
gas density, pressure, solar wind features and others are expected
to vary in these larger regions because the static and dynamic
conditions of matter and radiation are different. Thus, measurements
of the $Voyager$ $Probes$ refer indeed to the very skinny region
around a G star, the $Sun$,  and extrapolations of these
measurements to the huge and variegated interstellar space might be
insecure. \rm}} , quiescent electrons of the indifferentiated
interstellar medium are accelerated and decelerated with multiple
gains and losses of kinetic energies in a wide range of values. For
example, a local electric field of $10^{-6}$ $V$/$m$ acting in an
unshielded  milliparsec region ($3 \times 10^{13}$ $m$) imparts
$100$ $MeV$ of kinetic energies to charged particles. This implies
that the neutralization charge of quiescent electrons most likely
occurs not at thermal or suprathermal energies ( $>$ $70$ $eV$), in
the $keV$ band, but at higher energies due to the ubiquitous
presence of local electrostatic fields.

\section{The positive electric charge stored in the solar cavity}

\quad  The intensity of cosmic nuclei versus time during the
magnetic solar cycle of 22 years enables to infer the intensity of
cosmic nuclei in the outer space surrounding the solar cavity. The
energy spectra of the cosmic radiation are called $LIS$ ($Local$
$Interstellar$ $Spectra$) or demodulated spectra, or, eventually,
energy spectra of the very local interstellar medium ($VLIS$) as
hinted earlier.

\quad The most recent advance on the measurements of the electric
currents $I_{sw}$ and $I_{ds}$ have been made by the $V1$ and $V2$
Voyager detectors which directly determine cosmic-ray fluxes in the
very local interstellar medium. The energy spectra of cosmic proton
and electron are shown in fig. 4. The electron spectrum from radio
data \cite{peterson99} in the interval $400$ $MeV$ to $1$ $GeV$ (blue squares)
shown in fig. 4 intrinsically constitutes the demodulated spectrum
and, in this particular case, appropriately termed $LIS$ spectrum.

\quad Cosmic protons  in the range $3$ $MeV$-$10$ $GeV$ entering the
solar cavity have a flux of $15$$037$ part/$m^2$ $s$ $sr$ according
to the observed spectrum shown in figure 4 and 5. Due to the solar
modulation only a fraction $F_p$ of these protons contributes to the
current $I_{sw}$.  The fraction $F_p$ depends on the energy and it
is intended to be the average value in many solar cycles of 22
years. The fraction $F_p$ is shown in figure 6 according to
observations collected in the last 70 years. The proton fraction
$F_p$ is called unmodulated proton fraction. The positive charge
lost by cosmic protons in the volume $V_{sc}$ due to the flux $J_p$
is : $q$ $A_{sc}$ $\int ^ {E_{2}}_ {E_{1}} { J_{p}}$ $F_p$ $dE $  =
$3.92 \times 10^{12}$ $C$/$s$ where $q$ = $1.602 \times 10^{-19}$,
$A_{sc}$ = $2.2262 \times 10^{27}$, $E_1$ = $3$ $MeV$ and $E_2$ =
$20$ $GeV$. Details of the calculation are given in the $Appendix$
$B$.

\quad Cosmic electrons in the same energy range $3$ $MeV$-$20$ $GeV$
have a flux of $9$$219$ part/$m^2$ $s$ $sr$ according to the
observed spectrum shown in figure 4 and 5. The unmodulated electron
fraction, $F_e$,  versus energy is given in figure 6. The negative
charge entering the volume $V_{sc}$ corresponding to this flux is
$3.28 \times 10^{12}$ $C$/$s$. Therefore, the global electric charge
balance of proton and electron in the range $3$ $MeV$-$10$ $GeV$ is
 +  $0.64 \times 10^{12}$ $C$/$s$.

\quad Proton and electron spectra shown in figures 4 and 5
intercross around $60$ $MeV$. This energy represent a critical
divide : above $60$ $MeV$ the total electric charge entering the
solar cavity via cosmic rays is always positive while below $60$
$MeV$ is always negative. In the range $3$-$60$ $MeV$ the global
input charge of proton and electron is negative, \quad - $5 230$
$C$/$s$ while that above $60$ $MeV$ is positive,  being \quad  + $10
741$ $C$/$s$.

\quad In the global charge balance across the solar cavity positrons
and antiprotons at low energies have to be included. In the critical
range $1$ $MeV$ to $200$ $MeV$ no measurements of these
antiparticles in the interstellar medium are available. Thus, only
tentative extrapolations of the modulated spectra observed at Earth
might be used.

\begin{figure}
\begin{center} 
\includegraphics[width=0.8\textwidth]{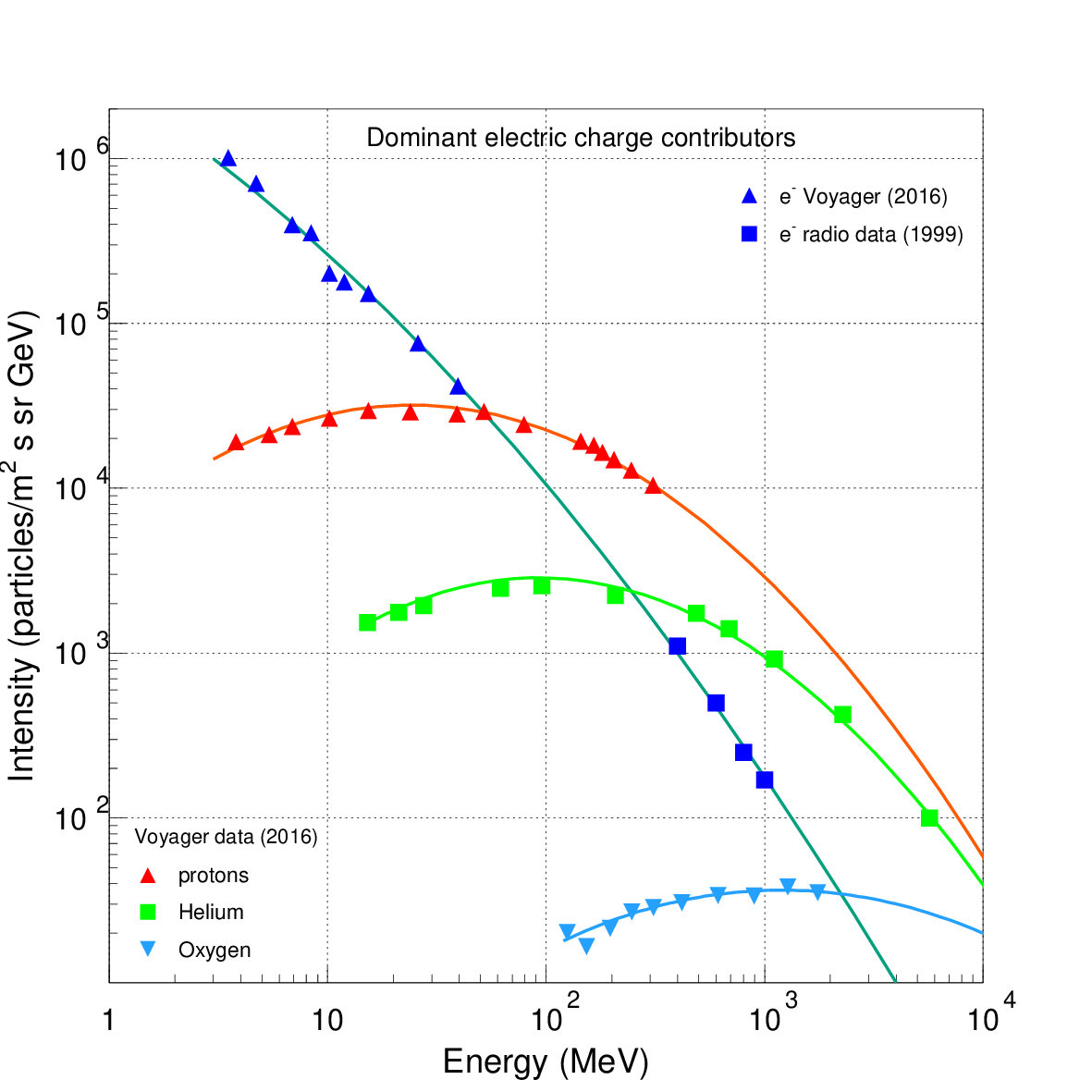}
\caption{The total electric charge permanently entering
the solar cavity comes from four dominant particles :  electrons,
protons, Helium and Oxygen. The four spectra shown in the figure are
the local interstellar energy spectra or demodulated spectra,
namely, those purified  by the solar modulation effect. These
spectra have been measured by the Voyager Probes V1 and V2 during
many years of observation.  At low energies, below $100$ MeV, solar
modulation at Earth can decrease the demodulated intensity by more
than an order of magnitude. \label{fig:111}}
\end{center}
\end{figure}

\quad The antiproton flux in the energy range $1$ $MeV$ to $200$
$MeV$ remains unmeasured. Just above $200$ $MeV$ the $\overline
p$/$p$ flux ratio is about $10^{-5}$ \cite{abe12} reaching a stable plateau
of  about $2 \times 10^{-4}$ above $2$ $GeV$. Loose upper limits of
about $10^{-1}$ at the maximum explored energies of $1$-$10$ $TeV$
have been reported \cite{bartoli12} using the standard Moon shadowing technique.
As antiprotons are secondary particles generated by interactions of
cosmic nuclei in the $Galaxy$ and no obvious sources are known,  the
negative charge from antiprotons entering the solar cavity is
negligible compared to that of low-energy cosmic electrons.

\quad Cosmic positrons are secondary particles generated by
interactions of cosmic nuclei in the $Galaxy$, but unlike
antiprotons,  Galactic quiescent positron sources do exist located
in stars and supernovae remnants. These sources are the radioactive
nuclides $^{44}Ti$, $^{56}Ni$ and $^{26}Al$ which yield positrons in
the $MeV$ range while decaying.

\begin{figure}
\begin{center} 
\includegraphics[width=0.8\textwidth]{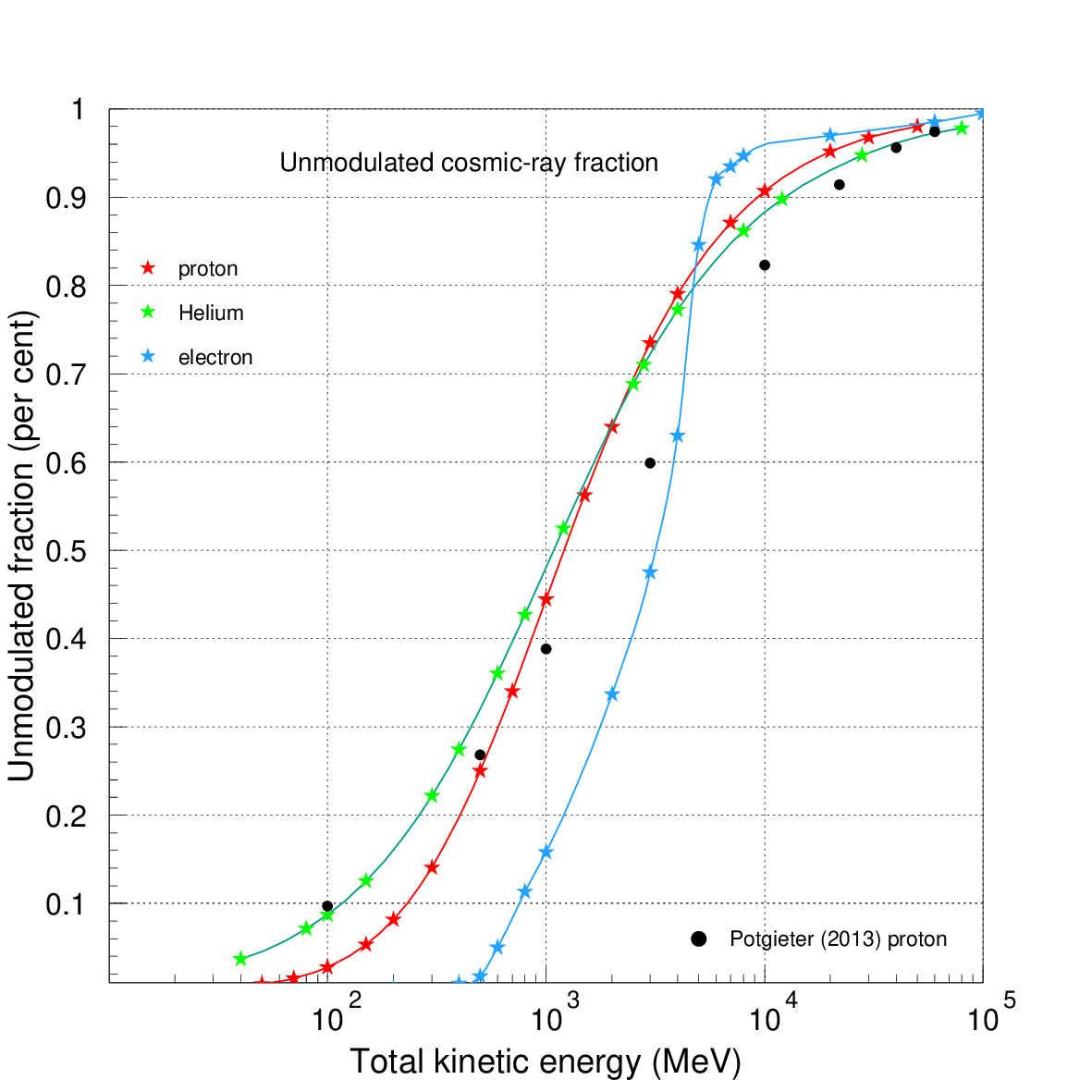}
\caption{Fractions of unmodulated intensities versus total
kinetic energy for electron, proton and Helium in the solar cavity.
Black dots are just an example of estimates of unmodulated proton
fractions \cite{potgieter13} that might serve for illustrative comparison. \label{fig:113}}
\end{center}
\end{figure}

\quad Important measurements of positron flux below $100$ $MeV$ at
Earth were performed in 1968 by the $IMP$-$8$ detector \cite{beuermann69} and in
2018 during the positive magnetic polarity of the $Sun$ cycle by the
AESOP-Lite detector in the energy band $20$ $MeV$ to $1$ $GeV$ \cite{mechbal20}.
Data points of the highest rates of these two experiments are shown
in fig. 4. Positron electric charge, in principle, might challenge
the dominance of negative electric charge carried by electrons in
the range $1$ $MeV$-$100$ $MeV$ (see fig. 4).

\quad Similarly to equation (1) the charge of the positive positron
current, $Q_{pos}$,  is computed by : $I_{pos}$ $\equiv$ $q$
$A_{sc}$ $\int $ $J_{pos}$ $dE$.

\quad  In principle low energy positrons from radioactive decays of
$^{44}Ti$, $^{56}Ni$ and $^{26}Al$ might be trapped in local
electric fields debouching in flux spikes at very low energies in
the band 1 $MeV$ to 50 $MeV$ thereby affecting the neutralization of
the positive charge $Q_{sw}$. Should the positron current $J_{pos}$
be comparable to the migration current $I_e$,  the postulated
neutralization around the $Sun$ becomes questionable.

\section{Electric charges stored by stars are consistent with the observed stellar magnetic fields}

\quad From the previous analysis follows that a current of negative
charges of quiescent electrons $I_e$ called $migration$ $current$ in
the work \cite{cod20ubibook} and depicted in fig. 1 has to enter stellar cavities of
any stars to neutralize the positively charged currents, \quad
$I_{sw}$ + $I_{ds}$ \quad certainly deposited by cosmic rays.

\quad  Moving charges generate magnetic fields of definite strengths
and shapes. As stars retain positive electric charge and move at
high velocities,  both rotating and translating, smooth magnetic
fields sprout everywhere. Here the focus is on the fact that
measurements of magnetic field strengths of various stellar
categories and $Sun$ are consistent with rotating electric charges
in the range $10^{19}$-$ 10^{20}$ $C$ with typical rotational
periods $0.1$-$50$ $days$ \cite{vidotto14}.

\quad The $Sun$ surface receives from the cosmic radiation the
positive charge per second $I_{ds}$ = $1.14 \times 10^{3}$ $C$/$s$
computed by $J$ = a/$E^{\gamma}$ with a = 28656 part/$m^2$ $s$ $sr$
$GeV^{1.67}$ \cite{cod20ubibook}, $\gamma$ = $2.67$ \cite{cod20ubibook}, a threshold energy $E_1$ =
$5$ $GeV$, $E_2$ arbitrarily large being not influential, $\overline q$ = $1.2 \times 10^{-19}$ $C$ \cite{cod15icrc} and $4$ $\pi$ $R_{s}^2$ = $6.2262
\times 10^{18}$ $m^2$ the $Sun$ collecting area. The threshold
energy of $5$ $GeV$ seems in the correct range taking into account
solar modulation. For example, a threshold of 10 $GeV$ would result
in $I_{ds}$  = $3.6 \times 10^{3}$ $C$/$s$.

\quad As particle density in the $Sun$ is  close to $8.3 \times
10^{23}$ $particles$/$cm^3$ from 1.4 $g$/$cm^3$/$1.67 \times
10^{-24}$ $g$, being exceedingly high relative to the electron
density of the migration current, the neutralization of the charge
$Q_{ds}$ occurs $in$ $situ$, i.e. cosmic nuclei extinguished within
solar photosphere absorb quiescent electrons from local solar
materials. If no charge neutralization from external sources takes
place, the requirement of charge conservation within the $Sun$
radius $R_{s}$ would yield an unlimited accumulation of positive
charge as electrons from the $migration$ $current$ from the local
interstellar medium are thoroughly absorbed by the ionized atoms of
the solar wind and cannot reach the $Sun$ main body which resides
within the tiny radius of $R_{s}$ = $6.98 \times 10^5$ $km$ face to
$R_{sc}$ of $1.331 \times 10^{13}$ $km$.

\quad For example, in one billion years ($\delta$$t$ = $3.155 \times
10^{16}$ $s$) assuming no charge neutralization from the migration
current $I_e$ , the accumulated charge in the $Sun$ would amount to
$I_{ds}$$\delta$$t$ = $3.61 \times 10^{19}$ $C$. Notice further that
prestellar materials do certainly store positive electric charges
deposited by cosmic rays as argued elsewhere (see $Segments$ $11.2$
and $12.4$ of ref.\cite{cod20ubibook}). This pristine charge deposition occurring in
the nascent $Sun$ certainly adds to the charge $I_{ds}$$\delta$$t$
computed above. With no charge neutralization, ineluctably, magnetic
field intensities in stars have to augment with time until charge
densities saturate the hosting and absorbing structures eventually
enabling new discharge channels.

\quad Global magnetic fields resulting from spinning charges can be
estimated by assuming a simple loop of current $i$ of radius $R$.
The order of magnitude of the field strength is, \quad $B$ = $\mu_0$
$i$/ ($2$$\pi$ $R$) \quad where $R$ is the radius of the star.
Measured star rotation periods range from $0.1$ to $200$ days (see
data of stellar rotational periods in $Table$ $3$ of ref.
\cite{vidotto14}19-vidotto; also fig. 6 in Reiners). For example, the $Sun$ rotation
period is $27.5$ days, $2.376 \times 10^{6}$ $s$ so that, after one
billion year,  $i$ = $Q_{ds}$/$T$ = $1.519 \times 10^{13}$ $A$. The
magnetic field strength of the rotating charge $I_{ds}$$\delta$$t$
from the formula above is : $B$ = $2.0 \times 10^{-7}$ $\times 1.519
\times 10^{13}$ $A$/$6.98 \times 10^{8}$ $m$ = $42 \times 10^{-4}$
$T$ = $42$ $G$. In the quiet $Sun$ the overall magnetic field
strength is in this range \cite{sanchez11} and, similarly,  in
other stellar categories \cite{plachinda99}.

\quad Perhaps the best measurements of magnetic field intensities in
stars are those in binary systems. Magnetic flux conservation is
invoked during stellar explosions in binary systems. Such explosions
yield neutron stars. The typical radius of collapsed stars is about
$10^{6}$ $km$ while that of a neutron star about $10$ $km$. In a
spherical geometry the ratio of the areas is $10^{10}$ which is the
enhancement factor of the magnetic field. Magnetic field strengths
measured in neutron stars are in the range $10^{11}$-$10^{12}$ $G$
and, consequently, with a reduction factor of $10^{10}$, those in
collapsing parent stars in binary systems are in the range
$10$-$100$ $G$.

\quad These magnetic field strengths are also measured, order of
magnitude, in the quiet Sun \cite{sanchez11}, in $F$, $G$, $K$ and
$M$ stars \cite{plachinda99} and $O$ and $B$ stars
\cite{kholtygin17}. A sample of magnetic field
strengths measured in $O$ and $B$ stars conform to a lognormal
distribution with an average value of 338 $G$. 

\quad  It is worth recalling that magnetic field strengths measured
in the spots, flares and energetic outbursts from the $Sun$ surface
and other star surfaces are approximately $1000$-$5000$ $G$, some 2
to 3 orders of magnitude higher than $10$-$100$ $G$. As magnetic
field intensities in stars in quiet conditions unambiguously differ
from those measured in magnetic spots, the nexus between rotating
charges and magnetic field intensities is highly suggested or
tentatively demonstrated.

\quad In order to explain magnetic fields in stars, aside from
spinning positive charges  $Q_{ds}$, other mechanisms have been
proposed such as the dynamo theory quite recurrent in the
literature.

\quad Neutron stars result from the explosion of massive stars which
have magnetic field strengths of some $\mu$$G$ believed to be
originated by the classical dynamo mechanism. As magnetic flux
during explosion conserves, field strengths in neutron stars have
enhanced magnetic fields  by a rough factor of $10^{10}$, as
previously noted. Do magnetic fields in neutron stars also originate
via dynamo mechanism ?  This mechanism seems inapplicable to neutron
stars which hardly might host the large convective cells of normal
stars due to the small sizes of about 10 $km$. Conceivably, rotating
positive charges appear to be a more natural explanation of both
magnetic fields in progenitor star and neutron star.

\section{Compendium  and Conclusion}
\quad It is an assessed fact that cosmic rays arriving at Earth are
arrested in the solar wind thereby depositing a positive electric
charge at all the energies above $60$ $MeV$ (see fig. 4) according
to the Voyager data. The positive electric charge per second
$I_{sw}$ deposited by cosmic-ray nuclei (proton and helium) in the
solar cavity of nominal radius $R_{sc}$ = $1.331 \times 10^{13}$ $m$
\cite{cod22hoewelebook,stone05} has been calculated in $Section$ $5$ and $Appendix$ $B$ of
this work and it amounts to $I_{sw}$ = $3.83 \times 10^{12}$
$C$/$s$. The estimated current $I_{sw}$ is likely to be correct
within a factor of 2.

\quad This charge  has to be neutralized in order to avoid electric
fields of very high intensity within the solar wind volume
\samepage{\footnote{\it The electric field $E_{sw}$ generated by the
permanent electric charge residing in the solar cavity is immersed
in a bath of moving ions, namely,  the solar wind,  which tends to
shield and, consequently to obscure any electrostatic effects. In
spite of that, during transient phenomena within the solar wind
itself or perturbations originated outside the solar wind volume,
shielding may be inefficient or absent. In this case electron
acceleration and nucleus acceleration within the solar wind have to
manifest. Indeed, proton and Helium acceleration in the range
0.1-$10$ $MeV$ in the interplanetary medium has been observed  long
time ago \cite{mcdonald76}; it was unpredicted and erroneously attributed to non
electric acceleration processes. A population of energetic electrons
in the solar wind in the interval $2$-$100$ $keV$ during quiet time
conditions has been also reported and its acceleration is not in the
Sun \cite{lin98} but in the interplanetary space. It is worth noting that
quite time conditions indicate absence of $CIR$ compressions,
absence of shocks in the medium  and no sudden ambient alterations.
\rm}} and its environment. Sources of electric charge within the
solar cavity in a finite, specified volume, for example Jovian
electrons \cite{lheureux76,moses87}, conserve electric charge and, accordingly, cannot
neutralize $I_{sw}$ over adequate, long time intervals. It follows
that the electric charge has to come from the exterior of the solar
cavity as qualitatively depicted in fig. 1 by blue arrows pointing
inward representing entrant negative charge (electrons). The motion
of these low energy electrons and their inherent electric currents
in the whole Galaxy, gives rise to the $migration$ $current$ denoted
by $I_e$ as asserted in this work and motivated elsewhere \cite{cod20ubibook,cod22ubi}. The
pristine spatial origin of the $migration$ $current$ $I_e$ is in all
cosmic-ray sources in the Galaxy as amply debated in ref. \cite{cod20ubibook}.

\quad The neutralization of the positive charge in the arbitrary
time interval $\delta$$t$ within the solar cavity of nominal radius
of $89$ $AU$ \cite{cod22hoewelebook,stone05} $Q_{sw}$ = $I_{sw}$ $\delta$$t$  requires an
equal amount of negative charge designated here by $Q^-_{ne}$ =
$I^-_{ne}$ $\delta$$t$ to be established by measurements. Ideally,
in steady state flows \quad $Q^-_{ne}$ $\delta$$t$ = $I_e$
$\delta$$t$ \quad where $I_e$ is the migration current based on a
logical inference amply debated in the work \cite{cod20ubibook},  as mentioned
earlier. The notion of $Q^-_{ne}$ derives from the necessity to
neutralize the electric charge deposited by cosmic nuclei
extinguished in the Galaxy. The Sun offers a unique opportunity to
measure the electric charge balance of cosmic rays around one star
and, hence, to observe one $migration$ $current$ around a single $G$
star and not the $global$ $migration$ $current$ of the $10^{11}$
Galactic stars which globally generate the Galactic magnetic field
(see ref. \cite{cod20ubibook}; $chapters$ 14 and 15).

\quad It is an impressive and notable fact that the negative current
$I^-_{ne}$ has been detected by the Voyager instruments $V1$ and
$V2$ by measuring huge rates of energetic electrons in the range $3$
to $60$ $MeV$ (see fig. 4) while exiting the solar cavity beyond the
shell region denoted $Heliosheath$ \cite{stone13}.

\quad The negative electric charge per second in the range $3$-$60$
$MeV$ entering the solar cavity $I^-_{ne}$ is - $1.86 \times
10^{12}$ $C$/$s$ derived in $Section$ $5$ and $Appendix$ $B$ from
the cosmic-ray electron spectra measured by the Voyager detectors.
This negative charge surprisingly counterbalances the positive
electric charge per second deposited by cosmic rays in the solar
cavity, $I_{sw}$ + $I_{ds}$, of $3.83 \times 10^{12}$ $C$/$s$ in the
range $60$ $MeV$-$20$ $GeV$. Within the accuracy of the calculation,
the missing negative charge labeled  here $N_-$ necessary for a
perfect charge neutrality, namely, \quad - $1.86 \times 10^{12}$ +
$N_-$ + $3.83 \times 10^{12}$ = 0  \quad is  - $1.97 \times 10^{12}$
$C$/$s$. This charge amount is not incompatible with the electron
spectra in the range $100$ $keV$-$3$ $MeV$ constrained by the
Pioneer data [26] and low frequency radio data \cite{peterson99} as speculated by
others in the solar modulation arena (see, for example, ref. \cite{langner01}
for a plausible electron spectrum in this unexplored energy range).

\quad In some respects the postulated dominance of cosmic electrons
over nuclei below $60$ $MeV$ is both impressive and surprising
\samepage{\footnote{\it The surprise  has been vented by those who
made the measurements like William R.  Webber of the Las Cruces
University, New Mexico, a colleague of balloon experiments of bygone
days (see for example ref. \cite{golden94,brunetti96}). In fact, he asserts \cite{webbervilla18}: " the
$LIS$ ratio of the electron to $H$ nuclei intensities in $MeV$,
e/H(E), measured up to $60$ $MeV$ by $V1$ outside the
heliosphere...At 2 $MeV$ this e/H(E) ratio is 100 (yes $100$ !)
decreasing to $1.10$ at $60$ $MeV$. " \rm}} because cosmic rays
above solar modulation energies, i.e. $10$ $GeV$- $3$ $TeV$, exhibit
the opposite trend, namely, fluxes of protons and heavier nuclei
dominate those of cosmic-ray electrons. The energy of $3$ $TeV$
above is the maximum measured cosmic-ray electron energy to day
(2025).

\quad In the census of electric charges within the finite volume
$V_{sc}$ and arbitrary time span $\delta$$t$,  the charge absorbed
inside the $Sun$ photosphere within $R_s$, namely $I_{ds}$
$\delta$$t$,  is negligible but it appears quite adequate to
generate stellar magnetic fields as loomed out in $Section$ $6$.

\quad Along the same logical framework of this calculation it
emerges that rotating electric charges stored by stars due to the
charge deposition of cosmic rays generate magnetic field strengths
in the range $10$-$100$ $G$ with dipolar geometry which are in
accord, order of magnitude, with spectropolarimetric Zeeman data of
$O$ stars and radio observations of magnetic fields of neutron star
progenitors as debated in $Section$ $6$. The agreement of computed
and observed magnetic field intensities of about $10$-$100$ $G$
further supports the context of this work.

\section*{Appendix A\\The novel scientific context in Cosmic Ray Physics}

\par\parskip=0.5truecm
\quad According to the works \cite{cod20ubibook,cod22ubi} the  $Milky$ $Way$ $Galaxy$ has a
pervasive and stable electrostatic field designated by $\vec E_g$ (g
for galactic) generated by the motion of positively and negatively
charged particles of the cosmic radiation. Cosmic rays are
accelerated by the electrostatic field which performs the
acceleration from quiescent energies up to the maximum energies of
$2.48 \times 10^{21}$ $eV$ of cosmic Uranium nuclei \cite{codino17}. In
restricted regions of the $Galaxy$ the field $\vec E_g$ is shielded
by ionized materials as, for instance, the region occupied by the
solar wind in the solar system.

\quad Facts supporting the existence of the $Galactic$ electrostatic
field discussed in \cite{cod20ubibook} are : \quad (1)  the constant spectral index
of the overall cosmic-ray energy spectrum comprised between
 2.64-2.68 up to energies of $2.8 \times 10^{19}$ $eV$ ;
  \quad   (2) the maximum energies of Galactic protons of
$2.8 \times 10^{19}$ $eV$;  \quad (3) the chemical composition of
the cosmic radiation above the energy of $2.8 \times 10^{19}$ $eV$
which has to consist only of heavy nuclei;  \quad (4) the
 intensity, orientation and direction of the Galactic magnetic
 field;  \quad (5) the absence of correlation of the arrival directions of
 ultra-high-energy cosmic rays in the interval $10^{18}$-
 $10^{20}$ $eV$ with powerful radio galaxies in the
 nearby universe.
 As cosmic rays
 are Galactic up to $2.48 \times 10^{21}$ $eV$, the
 absence of correlation is a natural, obvious in the context of the novel
picture of the cosmic radiation reported in the works \cite{cod20ubibook,cod22ubi,cod22hoewelebook}.

 \quad The calculated features (1), (2), (3), (4) and (5)
 impressively agree with the experimental
 data as debated and highlighted in the aforementioned research
 book \cite{cod20ubibook}.
 By contrast, results and  predictions recurrent in the past and present
 literature of the traditional theories of the cosmic radiation
 are severely inconsistent with
 the facts (1)-(5). For example, the bulk motion of cosmic rays
 in the erroneous traditional theories is
 thoroughly disconnected from the geometrical pattern of the
 Galactic magnetic field.

\quad Probably, the most outstanding result in terms of accord
between data and calculation reported in \cite{cod20ubibook}, designated fact (4)
above, is the exact account of the regular magnetic field of the
$Milky$ $Way$ $Galaxy$ with its intensity of $1$-$5$ $\mu$$G$ and
its variegated and unique geometrical pattern in all the Galactic
volume and beyond. The remarkable accord with the optical and mostly
radio data is described in $Chapters$ $14$ and $15$ of ref. \cite{cod20ubibook}.

\quad  Presently a conspicuous fraction of scientific community
believes that cosmic rays are accelerated in supernova remnants up
to $10^{15}$-$10^{16}$ $eV$ by a mechanism called diffusive shock
acceleration. Numerous and variegated experimental data disagree
with this credence. These inconsistencies are presented and debated
in the Appendix $C$ of ref. \cite{cod20ubibook} and in the research book \cite{cod13progbook} :
$Progress$ $and$ $Prejudice$ $in$ $Cosmic$ $Ray$ $Physics$ $until$
$2006$.

\section*{Appendix B\\ The modulated spectra of electron, proton and Helium}

\par\parskip=0.5truecm
\quad   In $Section$ $4$ to preserve a simple  calculation scheme of
the electric charge balance in the solar cavity, namely,  $I_{sw}$ +
$I_{ds}$,  only electrons and protons have been considered.  Here
two necessary extensions of the calculation : (1) the effect of the
solar modulation on the energy spectra ; (2) the inclusion of Helium
in the charge budget, the third major charge source in the solar
cavern. The next contributing heavier nucleus is Oxygen. Its charge
input, according to the data shown in figure 4,  is less than 8 per
cent of the charge input of proton and Helium and therefore
negligible for the aim of this work.

\quad The analytical representation of the demodulated spectrum,
$J_{de}$,  anchored to the observational data of Voyager Probes \cite{beuermann69}
and radio data \cite{peterson99} allow a reliable evaluation of the $modulation$
$factors$. These factors quantify  the amount of the observed flux
of energy $T$ (total kinetic energy) arrested in the solar cavity,
namely, $F$ = $J_{de}$/$J$ where J is the measured terrestrial flux
or that in another location within $1.331 \times 10^{13}$ $m$. The
modulation region goes from the
 Sun photosphere
up to 112 AU \cite{stone05,richa08}.

\begin{figure}
\begin{center} 
\includegraphics[width=0.8\textwidth]{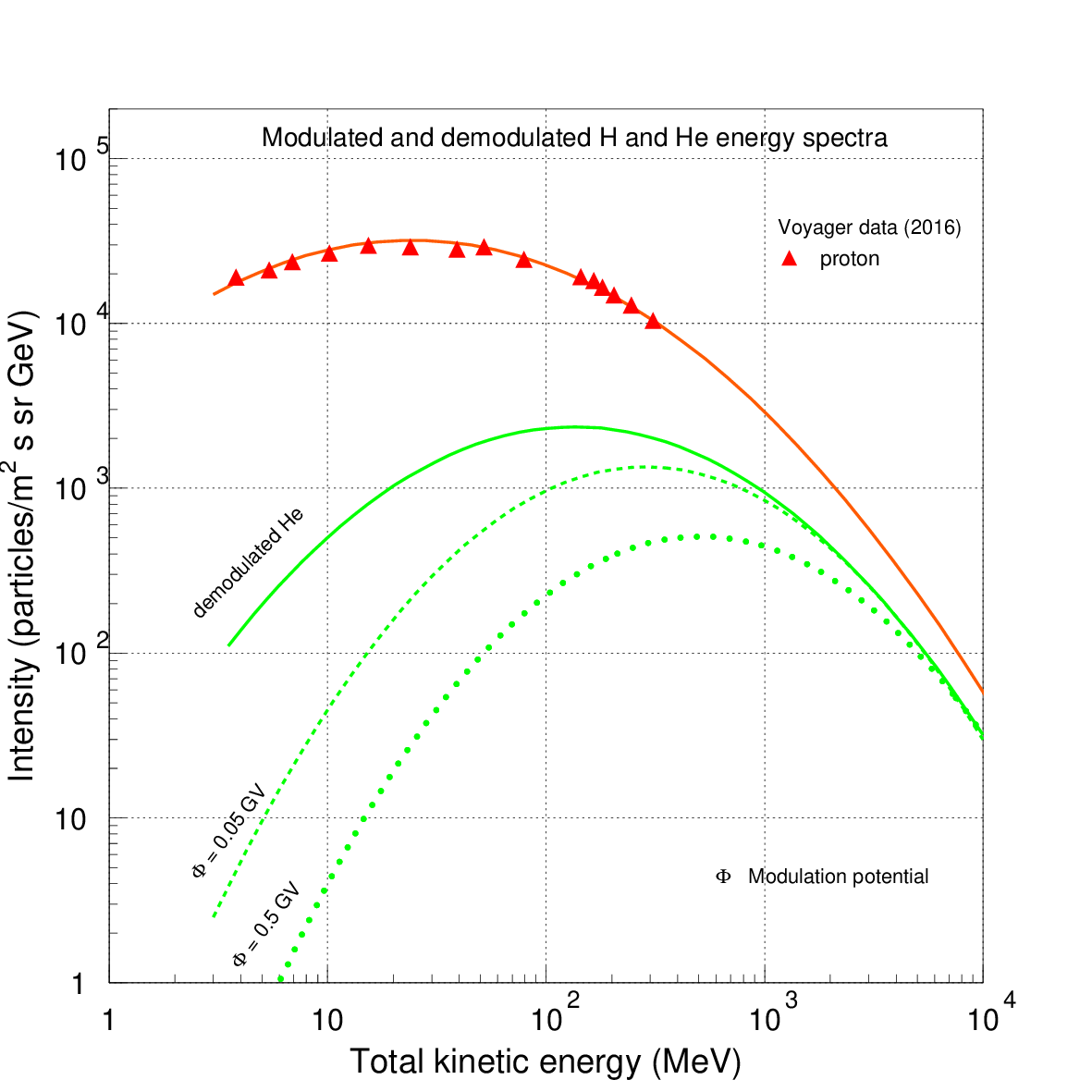}
\caption{Examples of solar modulation for cosmic-ray
Helium with two arbitrary values of the modulation potential, $\phi$
= $0.05$ and $0.5$ $G$$V$.  The interstellar Helium spectrum in the
range $3$ $MeV$-$10$ $GeV$ is also shown called, more appropriately,
demodulated Helium. Analytical expressions for the He spectra are
from ref. \cite{codino24}. To preserve simplicity demodulated Helium data are not
shown; they are in fig. 5. \label{fig:114}}
\end{center}
\end{figure}

\quad Since half a century it is known that the intensity variations
of proton and Helium registered on Earth in the range of $50$-$500$
$MeV$,  between minimum and maximum modulating conditions,  span a
factor of about $5$-$6$. This basic fact remains true to day after
exploring 4 solar cycles. The spectra displayed in figure 7 are an
example of computed modulation of Helium according to a classical
parametrization \cite{webber13} adopted in this work.

\quad Electron and proton charge flows are given in $Section$ $5$.
Here are those of Helium.

\quad Helium in the range $12$ $MeV$-$20$ $GeV$ entering the solar
cavity have a flux of $3668.7$ part/$m^2$ $s$ $sr$ according to the
observed spectrum shown in figure 5. Due to the solar modulation
only a fraction  of Helium nuclei, $F_{He}$,  contributes to the
current $I_{sw}$. The fraction $F_{He}$ depends on the energy and it
is intended to be the average value in many solar cycles of 22
years. The fraction $F_{He}$ is shown in figure 6 according to
observations collected in the last 70 years.  The positive charge
lost by Helium in the volume $V_{sc}$ due to the flux $J_{He}$ is :
$2$  $q$ $A_{sc}$ $\int ^ {E_{2}}_ {E_{1}} { J_{He}}$ $F_{He}$ $dE $
= $2.61 \times 10^{12}$ $C$/$s$ where $q$ = $1.602 \times 10^{-19}$,
$A_{sc}$ = $2.2262 \times 10^{27}$, $E_1$ = $12$ $MeV$ and $E_2$ =
$20$ $GeV$. Thus, the He charge input is $39$ percent of that of the
proton.

\quad For comparison with proton and electron, in the range
$12$-$60$ $MeV$,  the global $He$ input charge is , \quad  $0.0735
\times 10^{12}$ $C$/$s$ while that in the range $60$ $MeV$ -$20$
$GeV$ is $1.26 \times 10^{12}$ $C$/$s$.

\quad The modulation factors shown in figure 7 are obtained by the
demodulated spectra (often called $LIS$ for Local Interstellar
Spectra) and measured spectra at Earth, or close to it. As the Earth
is located at 1 $AU$ additional electric charge deposition occurs
between the $Sun$ and the Earth. As the spherical volume enclosed by
the Earth radius to the $Sun$ is only a fraction of ${9.5 \times
10^{-5}}$ of that of the solar cavern, this charge deposition is
neglected in this work. Notice also that a severe conflict between
measured and computed radial cosmic-ray gradients emerged in the
inner heliosphere making uncertain plain extrapolations of charge
deposition based on the outer heliosphere data.

\quad  The dominant error source  in the evaluation of charge
$I_{sw}$ + $I_{ds}$ comes from the electron spectrum below $3$ $MeV$
which is unmeasured but highly constrained by radio data \cite{peterson99} and
Pioneer 10  observations \cite{lopate01} made at 70 AU for electron energies of
2-20 $MeV$. The limit of $3$ $MeV$ above is an intrinsic limitation
at low energy of the $V1$ and $V2$ Voyager instruments.

\section*{Appendix C\\ Failures of diffusion equation applied to cosmic rays}

\quad In principle the omission of the Galactic electrostatic field
in any calculation of the properties of Galactic cosmic rays has to
conduct to severe inconsistencies with  observational data.  Here
some conflicts between computed and observed cosmic-ray features are
mentioned. The computed features are obtained by  the $Diffusion$
$Equation$ of $Galactic$ $cosmic$ $rays$ used since more than  70
years. These conflicts are commonly registered in the literature.

\quad (1)  The Boron-to-Carbon flux ratio
 in the range $10$-$25$ $MeV$ is inconsistent with standard
 calculations ( see fig. 9 of ref. \cite{cummings16}).

\quad  (2)  According to the $V1$ Voyager Team \cite{cummings16} the $^2H$ and
$^3He$ isotope spectra in the range $5$-$50$ $MeV$/$u$ severely
disagree with those computed by the $GALPROP$ simulation code of
Galactic cosmic rays in spite of the fine tuning of simulation
parameters. In the energy band $10$-$100$ $MeV$/$u$ the observed
intensity of $^3He$ is higher than that computed by $GALPROP$ code
by more than an order of magnitude. The GALPROP code adopt the
diffusion equation tuned, in this case, at low energies around
0.005-20 $GeV$/$u$.

\quad  (3) The ionization rate of Hydrogen termed \quad $\zeta_{cr}$
($c$$r$ for cosmic rays) \quad computed with the reconstructed
cosmic-ray energy spectrum in Galactic environment via $Diffusion$
$Equation$ amounts to ($1.51$-$1.64$)$\times 10^{-17}$ $sec^{-1}$
\cite{cummings16}. It differs by more than an order of magnitude from the
ionization rate extracted from chemical reactions in the
interstellar medium, namely, $\zeta_{c}$ ($c$ for chemical) = $1.78
\times 10^{-16}$ $sec^{-1}$ [34]. The reconstructed energy spectrum
of cosmic rays in Galactic environment is derived from the diffusion
equation with its highly abstract and unrealistic scheme.

\quad  (4)  The diffusion equation of cosmic rays in the $Galaxy$
predicts a decreasing ratio of the antiproton-to-proton flux ratio
($\overline p$/$p$) above the energy of 2.6 $GeV$. On the contrary,
recent and past data in the interval $3$-$400$ $GeV$ show a flat
$\overline p$/$p$ ratio incompatible with calculations.

\quad Additional evidence of the inadequacy of current ideas to
describe low-energy cosmic rays, solar energetic particles and other
solar phenomena is elsewhere \cite{codino24}.

\end{document}